\documentstyle[12pt,titlepage]{article}

\begin{document}

\title{Global geometry optimization of clusters\\ using a growth
       strategy\\ optimized by a genetic algorithm}

\author{{\em Bernd Hartke\/}\\ \\
        Institut f\"ur Theoretische Chemie\\
        Universit\"at Stuttgart\\
        Pfaffenwaldring 55\\
        70569 Stuttgart, GERMANY\\
        e-mail: bernd.hartke@rus.uni-stuttgart.de}
\date{Chem.Phys.Lett., in print, 1995}

\maketitle

\begin{abstract}

A new strategy for global geometry optimization of clusters is presented.
Important features are a restriction of search space to favorable
nearest-neighbor distance ranges, a suitable cluster growth representation
with diminished correlations, and easy transferability of the results to
larger clusters. The strengths and possible limitations of the method are
demonstrated for Si$_{10}$ using an empirical potential.

\end{abstract}

\section{Introduction}

In recent years, many new methods have been developed for the problem of
cluster geometry optimization (see, e.g.,
Refs \cite{CP,JCP,Scheraga1,Scheraga2,Doll,Tsoo,Poteau,LeGrandMerz,JPC}), that
is, the search for the lowest energy configuration of a cluster of a given
number of atoms or molecules with a given interaction.
Also, attempts are being made to discover links between features in model
potentials and the resulting minimal energy structures \cite{Wales}.
Nevertheless, the following difficulties persist and are not fully solved
(or, partly, not even addressed) by the above methods:
\begin{enumerate}
   \item exponential increase of the number of local minima with cluster
         size \cite{Hoare};
   \item suitable representation of the problem (e.g., choice of coordinates
         or parameters to be optimized) \cite{LeGrandMerz,JPC,NFL};
   \item proper restriction of the search space \cite{JPC,Holland,Goldberg};
   \item high expense of ab initio potential energy surface calculation.
\end{enumerate}
Item (1) is well-known. Much less attention is paid to items (2) and (3),
although they are just as important as the choice of a good optimization
algorithm. Finally, the number and location of minima on a potential energy
surface depends critically on details of the potential energy function used
\cite{HoareMcInnes}. Therefore, ideally, geometry optimizations should be
performed at the ab initio level of theory; in that case, item (4) is of
paramount importance.

Recently, it has been proven that one cannot hope for discovery of a global
optimization method that performs well for every conceivable problem and
problem representation (``no free lunch theorem'' \cite{NFL}). Hence, one
should not try to tweak an existing optimization strategy to perform
acceptably for a given problem, but instead construct a new strategy based
upon prior information on the problem. In this Letter, I present GOALS
(global optimization by optimized assembly of local structure), a strategy
specifically designed to address problems (1)-(3) mentioned above, by the
following features:
\begin{enumerate}
  \item transfer of knowledge from small clusters to larger ones;
  \item no direct optimization of simple coordinates of individual atoms
        (which will always exhibit a high degree of correlation);
  \item all obviously nonsensical cluster structures are avoided by
        construction.
\end{enumerate}
Problem (4) will be dealt with in a subsequent paper.

One ingredient of the new method is a cluster growth strategy. Such growth
strategies have been used before (see, e.g., Refs \cite{Poteau,Vlachos} and
Refs cited therein, in particular also \cite{Northby}),
but only with rather crude ad-hoc methods of adding new atoms, followed by
local optimization. Here, the different ways of growth themselves are
subject to optimization.

It is important to emphasize that this Letter is a presentation of a new
method, {\em not\/} an investigation into particular cluster structures
or into the quality of a particular empirical potential.

This Letter is organized as follows: Section 2 contains the presentation of
the general idea and of a particular implementation; in section 3, this
implementation is applied to small silicon clusters; in section 4, limitations
and further work are discussed.

\section{Method}

\subsection{General idea}

The central idea of GOALS is to construct global energy minimum structures of
{\em large\/} clusters (of $N_L$ atoms) without explicitly subjecting them to
a global optimization strategy. Instead, structural information from globaly
optimized {\em smaller\/} clusters (of $N_c$ atoms) is used to build up
starting geometries for the large clusters wich are then only locally
optimized. At the same time, the problem of coordinates and proper search
space restriction is avoided by construction, in the following way:

``Cluster assembly agents'' are introduced: Each agent can act on a cluster of
$N$ atoms, for arbitrary values of $N$. It selects a particular site next to
the cluster (or inside the cluster) according to given criteria, and adds a
new atom such that a specific geometric arrangement is generated at this site,
for example, a face capping or a dangling atom.  The parameters to be
optimized are then simply the relative probabilities of these agents. Proper
design of these agents can approximate a one-to-one relationship between
agents and local structural features in a cluster. In this way, correlations
between parameters to be optimized are kept minimal.

Optimization of agent probabilities proceeds via a series of assemblies of a
cluster of $N_c$ atoms, starting from a small seed structure of $N_s$ atoms.
More atoms are added by choosing cluster assembly agents according to their
probabilities, at each stage $N_s \leq N < N_c$; the potential energy of the
resulting cluster of $N_c$ atoms is minimized by variation of the agent
probabilities.

After completion of this stage, the resulting set of optimized cluster
assembly agent probabilities can then be applied to generate good starting
guesses for traditional, local optimizations of {\em larger\/} clusters (of
$N_L$ atoms, with $N_L > N_c$). These local optimizations will arrive at the
global energy minimum structure(s) with unusually high probability, provided
the local structures present in the global minima geometries of larger
clusters are not very different from those found in the global minima
geometries of smaller ones.

\subsection{Preliminary implementation}

In this Letter, I present a particular, preliminary implementation of
GOALS. In order to keep the implementation simple, several approximations to
the general concept are introduced:

Each cluster assembly is done initially on fcc lattice sites only. The fcc
lattice was chosen simply because fcc and hcp are known to be the closest
packings of hard spheres in 3D, for $N\to\infty$ \cite{Conway} (the problem of
optimal cluster packings with $N<\infty$ is still unsolved, though).  Contrary
to the hcp lattice, the fcc lattice points are easily encodable via an oblique
coordinate system. Furthermore, many other lattices are related (in a formal
way) to the fcc lattice. Note, however, that the fcc lattice only serves as a
temporary template to establish neighborhood relationships.

As an approximation to the cluster assembly agents of section 2.1, I introduce
``numerical fcc-agents''.  They are {\em solely\/} characterized by the number
of nearest fcc-neighbors the atom will have that they are about to introduce.
For example, imagine a triangle of 3 atoms on neighboring fcc lattice sites.
The ``1-agent'' adds another atom such that this new atom will have exactly 1
nearest neighbor in the resulting 4-atom structure. That is, it introduces a
dangling atom at one corner of the triangle. Similarly, if we instead let the
``2-agent'' operate on the trianlge, we will get a flat or bent rhombus.
Action of the ``3-agent'' on the triangle generates a tetrahedron. Obviously,
none of the other agents (4-agent through 12-agent) is applicable to a 3-atom
triangle. Of course, these numerical fcc-agents are a very crude
approximation, but they have the following advantages: There are only twelve
possible agents, thus we cannot miss one inadvertently.  All twelve agents are
easy to code, no exercises in artificial intelligence or pattern recognition
are necessary. Nevertheless, the number of nearest neighbors is an essential
feature which is directly connected to sterical crowdedness vs.\ openness, and
to chains vs.\ flat structures vs.\ 3D ones. Disadvantages of this choice will
be discussed in section IV.

In order to arrive at realistic structures, independent of the fcc lattice,
each completed cluster is subjected to a traditional, local conjugate
gradient optimization \cite{NumRec}. The resulting minimized potential energy
is taken as characteristic for the whole cluster assembly that lead to the fcc
structure prior to the local optimization.

In order to destroy accidental symmetries present in the fcc-based geometries,
small random displacements are added to all coordinates before doing the local
optimization step. Both the assembly and the random displacements preclude a
one-to-one relationship between the parameters to be optimized and the cluster
geometries. Possible artifacts arising from this ambiguity are avoided by
taking the average potential energy from five different sequences [assembly +
random displacement + local optimization], for each set of agent probabilities.

The agent probabilities themselves can be optimized using any global
optimization method; here I am using the standard genetic algorithm (GA) as
described by Holland and Goldberg \cite{Holland,Goldberg} and used in Ref
\cite{JPC}. In a GA, the optimization problem is mapped onto an algorithm that
mimics natural evolution. In particular, the parameters to be optimized are
called genes, one particular instance of a complete parameter set is the
genome of an individuum, and many such individuals form a population. The cost
function measures the fitness of an individuum, which in turn determines how
many children this individuum will have in the next generation. Optimization
proceeds through mutation, exchange of information between individuals (via
genetic crossover), and preferential inheritance of favorable genes.  In the
present case, each individual in the GA consists of 12 values for the agent
probabilities; each of these values is encoded as a 5-bit binary number (this
is a compromise between short genes and a sufficiently fine grid). An elitist
strategy is employed (the best individuum of each generation is guaranteed to
survive unchanged into the next), since it seems to be performing slightly
better than the standard non-elitist version. It turns out that fairly small
GA's are sufficient in this context: 10 generations of 10 individuals each
already yield satisfactory results.

A separate program takes a given cluster assembly agent probability
distribution, builds new clusters of arbitrary size (again, on the fcc
lattice), and locally optimizes them (in cartesian coordinates, independent of
the fcc lattice sites). If this whole concept works, the probability to find
the global optimum for these new cluster sizes will be significantly larger
for agent probabilities optimized for smaller clusters than for random agent
probabilities or for completely random starting geometries.

To summarize, the final algorithm to be applied in the next section is this:
\begin{description}
   \item[stage 1:] optimize agent probabilities for an $N_c$-atom cluster by
      iterating the following steps:
      \begin{itemize}
         \item choose agent probability values
         \item starting from the $N_s$-atom seed, use these probabilities to
               build up 5 $N_c$-atom clusters
         \item locally optimize these 5 clusters; the resulting minimal
               energies are $E_i$, $i=1,\ldots,5$
         \item let $E_{av}=\frac{1}{5}\sum_{i=1}^{5}E_i$
      \end{itemize}
      and by minimizing $E_{av}$. Actually, this whole stage is done by the
      GA, but this is omitted for clarity.
   \item[stage 2:] construct optimized larger clusters by the following steps:
      \begin{itemize}
         \item use the optimized agent probabilities resulting from stage 1
               to build up clusters of $N_L$ atoms from the same $N_s$ seed
         \item locally optimize these clusters.
      \end{itemize}
\end{description}

\section{Exemplary application}

The first application of the GOALS strategy focuses on small silicon clusters
and employs the empirical potential developed by Bolding and Andersen
\cite{BoldAnd}. This potential is particularly good for small clusters, as
opposed to many others; and it is fairly complicated, hence effects of a
complexity similar to an ab initio potential can be expected. The objective
of this Letter is not an exhaustive exploration of this potential for large
$N$; therefore, this application is limited to clusters up to $N=10$, for
which the most important minimum structures are given in Ref \cite{BoldAnd}.

In converting from fcc structures to cartesian coordinates (required as input
for the local optimization on the Bolding-Andersen potential), I used the
Si-Si distance of bulk silicon as distance between nearest-neighbor fcc
lattice sites. This turns out to be a very reasonable choice, since most fcc
structures change only marginally upon optimization. This observation also
allays the suspicion that the choice of the fcc lattice could impede the
algorithm in this application, as silicon prefers the diamond lattice in the
bulk crystal.

As a first, trivial check of stage 1 of GOALS, I have taken a 3-atom seed (a
triangle on fcc lattice points), and optimized the cluster assembly agent
probabilities for a target cluster of $N=4$ atoms, i.e., just one more atom
is added. The global minimum structure for $N=4$ on the Bolding-Andersen
surface is a planar rhombus, hence the following outcome is to be expected
(cf.\ section 2.2):
The 2-agent should be clearly favored over
the 1-agent and the 3-agent. Obviously, all other agents are irrelevant
for this example. In fact, the best individuum after 10 generations shows
the following probability distribution: 0.19 for the 1-agent, 1.0 for the
2-agent, and 0.0 for the 3-agent (for simplicity, each single probability can
take on values between 0 and 1; the algorithm used does not require that all
probabilities sum to 1.0). This shows that the agent probability optimization
part of GOALS works as expected.

A more realistic example for the full GOALS strategy is the global geometry
optimization of Si$_{10}$. Fig.\ 1 summarizes the results (the numbers shown
are no full statistics, but merely particular runs; but I have checked that
they are representative):

Traditional, local optimizations of starting geometries with the coordinates
of all atoms chosen at random have practically no chance of finding the
global minimum at all. In 1000 runs, the global minimum at -39.2 eV was not
found even once; and the lowest structure found (-33 eV) was not even
close to it.

A direct GA-optimization of the cartesian coordinates of all atoms (not shown)
suffers from correlations between the coordinates, that is, the problem
representation is unsuitable. Hence, this approach has problems in finding
the global minimum structures of clusters with $N \geq 6$.

The next step in sophistication is generating starting geometries for local
optimization by the cluster assembly agent approach, but with random
probabilities for the agents (that is, using only stage 2 of section 2.2). In
such structures, all nearest neighbor distances are in a favorable range;
hence, this amounts to a massive but intelligent restriction of the search
space. Therefore, the chances for finding the global minimum are drastically
improved to 4\%.

Using agent probabilities optimized for smaller clusters (i.e., the full GOALS
strategy, stages 1 and 2 of section 2.2) leads to further improvement: If
agent probabilities optimized for $N=8$ are used for generating starting
geometries for optimization of the $N=10$ cluster, the chances for finding the
global minimum rise to 21\%.  In fact, the overall performance is even better
than this single number suggests: The chances of hitting one of the two lowest
stationary points are close to 40\%, and it is very hard to generate
unfavorable structures with energies higher than -32 eV.

If, however, agent probabilities optimized for $N=5$ are used, chances for
finding the global minimum for $N=10$ are with 6\% only marginally better
than with random probabilities. The reason for this is a structural transition
from planar geometries to 3D ones, which occurs between $N=5$ and $N=7$ on
the Bolding-Andersen surface. Such a drastic change is a peculiarity of the
small cluster range studied here; there are also structural transitions in
the higher-$N$ range \cite{VanDeWaal}, but they are less severe.

\section{Discussion}

It must be emphasized that the Si$_{10}$ example given is not trivial:
Lennard-Jones cluster studies have completely enumerated all minima up to
$N=13$ \cite{Hoare}. For $N=8$, only 8 minima are found, but 57(!) minima
for $N=10$. There is no complete enumeration of all local minima available for
the Bolding-Anderson potential, but it is likely that the situation is
similar.

The numerical fcc agents are only a crude approximation to true cluster
assembly agents envisaged in section 2.1. By construction, there are only 12
numerical fcc agents, and one can easily think of structures that are not
covered by them, for example C$_{60}$-like geometries.  Even more aggravating
is the fact that some agents do lead to unique structural features, but others
do not: For example, the 12-agent obviously has a singular meaning; similarly,
the 1-agent generates only dangling atoms or (linear or bent) chains. The
3-agent, however, can mean anything between a closed tetrahedron, a capped
triangular outside face, a reconstructed surface, or even a fully planar one.
This ambiguity is reflected in the results shown above: the probability for
this agent shows wide variations, even when others are already well converged.
Similarly, the 4-, 5-, and 6-agents can still lead to (planar or
reconstructed) 2D forms instead of 3D structures, albeit with less
probability. This is not in accord with the original idea of assembly agents:
Each agent should lead to a unique structural feature in the resulting
cluster; in this way, correlations between agent probabilities are minimized.
Therefore, one can conclude that there is presumably some significant
correlation left between the probabilities for the numerical agents.

The fact that the method works in spite of these shortcomings shows that
it also contains important, favorable features. As mentioned in the
introduction, these are:
\begin{itemize}
  \item suitable restriction of search space;
  \item diminished correlation between parameters to be optimized;
  \item flattening of the exponential-$N$-cliff of cluster geometry
        optimization, by easy transferability of cluster assembly agent
        probabilities from small $N$ to large $N$.
\end{itemize}
It can be argued that the possibility of unforeseen structural transitions
totally precludes the application of the GOALS strategy in the latter sense.
On the other hand, while the problem of structural transitions cannot be
denied, strategies like GOALS that transfer information from smaller clusters
or cluster subregions to larger clusters are the only way to weaken the full
impact of the exponential increase in complexity of the cluster geometry
optimization problem.

Further exploitation of the GOALS idea clearly calls for development of better
cluster assembly agents. Another aim of future work in this area is geometry
optimization on the ab initio level; to this end, methods for more effective
use of ab initio potential energy and gradient information are being
developed.

\newpage

\section*{Figure caption}

{\bf Figure 1:} Probability in \% of finding stationary Si$_{10}$ structures,
using several optimization methods. Figure 1b is a blow-up of the leftmost
part of Figure 1a. On the x-axis, the potential energy of the structures is
given in eV; for clarity, all structures within 1 eV are binned together. This
distorts the presentation somewhat; for example, the bin at -39 eV contains
only one structure, the global minimum. The results shown are from 100 local
optimizations of starting geometries generated by the following methods:
randomly chosen atomic coordinates (``random'', $\times$), fcc cluster assembly
using random agent probabilities (``random-fcc'', $\Box$), fcc cluster assembly
using agent probabilities optimized for Si$_5$ (``opt-5'', $+$), and fcc
cluster assembly using agent probabilities optimized for Si$_8$ (``opt-8'',
$\Diamond$). For further details, see text.

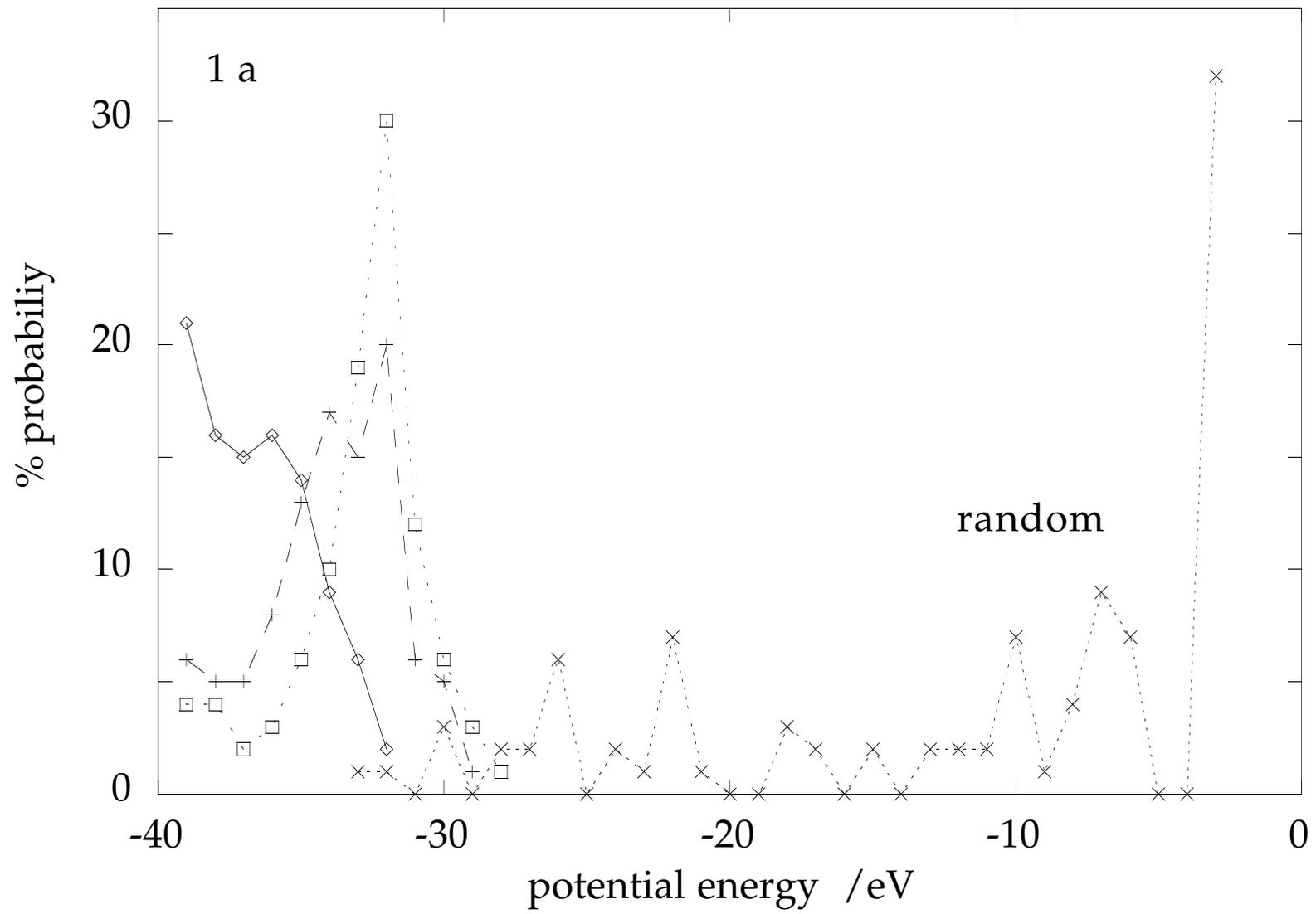
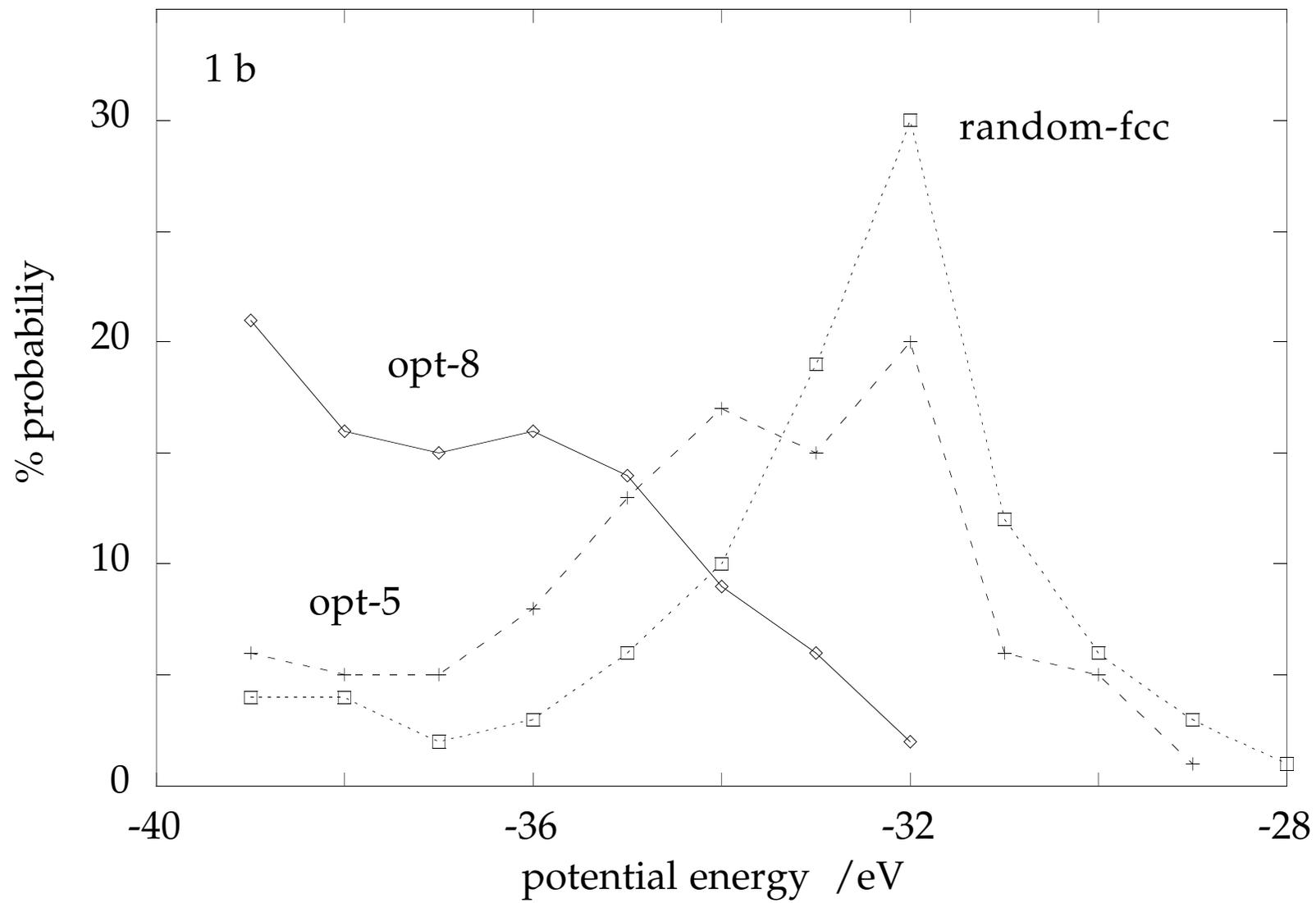
\end{document}